\pdfoutput=1
\documentclass[11pt]{article}

\usepackage[final]{acl}

\usepackage{times}
\usepackage{latexsym}

\usepackage[T1]{fontenc}
\usepackage[utf8]{inputenc}

\usepackage{microtype}

\usepackage{inconsolata}

\usepackage{graphicx}

\usepackage{amsmath}
\usepackage{multirow}
\usepackage{multicol}
\usepackage{booktabs}
\usepackage{amssymb}
\usepackage{url}

\makeatletter
\renewcommand{\@makefnmark}{}
\renewcommand{\@makefntext}[1]{#1}
\makeatother

\title{HSDreport: Heart Sound Diagnosis with Echocardiography Reports}

\author{
  \textbf{Zihan Zhao\textsuperscript{1,2*}},
  \textbf{Pingjie Wang\textsuperscript{1,2*}},
  \textbf{Liudan Zhao\textsuperscript{1,4*}},
  \textbf{Yuchen Yang\textsuperscript{2,3}},
\\
  \textbf{Ya Zhang\textsuperscript{1,2}},
  \textbf{Kun Sun\textsuperscript{4}},
  \textbf{Xin Sun\textsuperscript{4}},
  \textbf{Xin Zhou \textsuperscript{4}},
\\
  \textbf{Yu Wang\textsuperscript{1,2\dag}},
  \textbf{Yanfeng Wang\textsuperscript{1,2\dag}}
\\
  \textsuperscript{1}Shanghai Jiao Tong University,
  \textsuperscript{2}Shanghai AI Laboratory,
  \\
  \textsuperscript{3}University of Science and Technology of China,
  \\
  \textsuperscript{4}Xinhua Hospital Affiated to Shanghai Jiao Tong University School of Medicine
 \\
 \small{
    {\{zihanzhao,pingjiewang,sjdyszld,yuwangsjtu,wangyanfeng622\}@sjtu.edu.cn}
  }
  \thanks{\textsuperscript{*}Equal contribution}
  \thanks{\textsuperscript{\dag}Corresponding author}
}

\begin{document}
\maketitle

\begin{abstract}
%现有的心音诊断(HSD)任务多限定了几种疾病，将HSD任务设定为少量类别的heart sound 分类任务，在取得了一定的成果的前提下，并不完全符合医学实践，能够给医生提供的信息有限。
Heart sound auscultation holds significant importance in the diagnosis of congenital heart disease. However, existing methods for Heart Sound Diagnosis (HSD) tasks are predominantly limited to a few fixed categories, framing the HSD task as a rigid classification problem that does not fully align with medical practice and offers only limited information to physicians. Besides, such methods do not utilize echocardiography reports, the gold standard in the diagnosis of related diseases. To tackle this challenge, we introduce HSDreport, a new benchmark for HSD, which mandates the direct utilization of heart sounds obtained from auscultation to predict echocardiography reports. This benchmark aims to merge the convenience of auscultation with the comprehensive nature of echocardiography reports. First, we collect a new dataset for this benchmark, comprising 2,275 heart sound samples along with their corresponding reports. Subsequently, we develop a knowledge-aware query-based transformer to handle this task. The intent is to leverage the capabilities of medically pre-trained models and the internal knowledge of large language models (LLMs) to address the task's inherent complexity and variability, thereby enhancing the robustness and scientific validity of the method. Furthermore, our experimental results indicate that our method significantly outperforms traditional HSD approaches and existing multimodal LLMs in detecting key abnormalities in heart sounds.

%For this purpose, we have collected a new dataset named AuscultCardiography, which includes 2800 heart sound segments and their corresponding echocardiography reports. Since echocardiography reports are composed by physicians observing ultrasound images, and thus exhibit a modal gap with heart sounds,

%we propose a novel knowledge-aware query-based transformer for the new task, aiming to leverage relevant knowledge within the large language model (LLM) to handle diverse and complex echocardiography reports, thereby enhancing the robustness and scientific validity of the method. Furthermore, our experimental results indicate that our method significantly outperforms traditional audio-based classification approaches and existing multimodal LLM in detecting key abnormalities in heart sounds.

\end{abstract}
\section{Introduction}

%Congenital heart disease (CHD) is the most common congenital abnormality, with 13.3 million patients worldwide in 2019~\cite{roth2020global}. Between 150,000 and 200,000 newborns in China are diagnosed with CHD each year~\cite{of2022report}. Of them, around one-third will pass away in their first year and risk for mortality, morbidity and handicap will significantly raised in the surviving children if with delayed identification and treatment~\cite{pan2022trends}. Timely recognition of CHD is essential for prompt care and prevention of irreversible consequences.

%Cardiac auscultation remains a vital component of clinical medicine and has been widely used to screen for CHD due to its cost-effective. However, it heavily relies on clinical expertise and the acoustic hearing range of human ear~\cite{montinari2018history,mangione1997cardiac}. 所以心音听诊任务被提出了。但是当前的心音数据集多为二分类数据集，受限于二分类数据集和医学实践的差异，当前的数据集为实际的医疗诊断能够提供的参考信息有限。Echocardiography对于诊断CHD有着更全面的信息, 但 its accessibility in regions with limited healthcare resources is constrained by its time-consuming, high cost and stringent technical proficiency~\cite{wang2021automated}. 而if we can 结合听诊的便利性与echocardiography的全面性，将会enable the screening of a wider population并给医生提供更全面的参考. 

\begin{figure}[t]
    \centering
    \includegraphics[width=\linewidth]{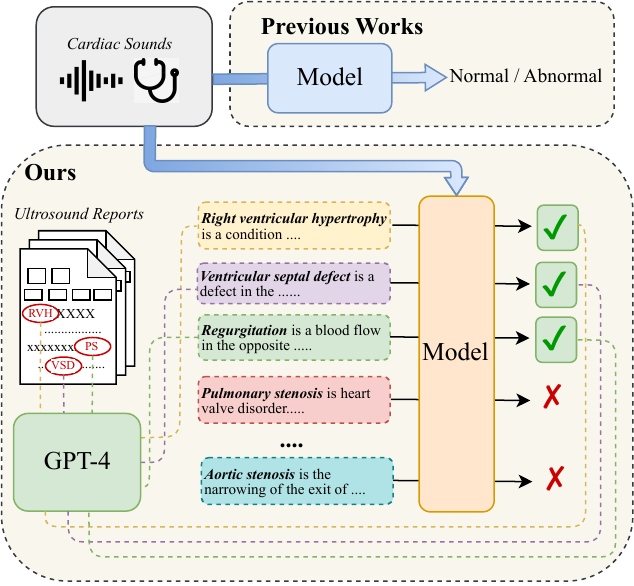}
    \caption{Comparison between previous works and ours. In previous works, heart sound diagnosis (HSD) was treated as a multi-class problem with two to five categories. Our new benchmark, starting from echocardiography reports, treats HSD as a twelve-category multi-label task. Furthermore, we have developed a knowledge-aware, query-based transformer approach that utilizes medical descriptions to address this novel benchmark.}
    % \label{fig:enter-label}
\end{figure}

Heart sound auscultation is an essential part of clinical medicine and is extensively utilized to screen for congenital heart disease (CHD) due to its cost-effectiveness. CHD is the most common congenital abnormality, with 13.3 million patients worldwide in 2019~\cite{roth2020global,of2022report}. For newborn patients, around one-third will pass away in their first year and risk for mortality~\cite{pan2022trends}. So massive promotion of heart sound auscultation is crucial. However, its effectiveness heavily depends on the clinician's expertise and the human ear's acoustic range~\cite{montinari2018history,mangione1997cardiac}. Consequently, the task of heart sound diagnosis (HSD) has been highlighted~\cite{clifford2016classification,oliveira2021circor,yaseen2018classification}.  

%However, current HSD datasets are predominantly limited to a few diseases, framing the HSD task as a classification of heart sounds into a small number of categories. While these datasets have achieved certain results, they offer limited reference information for actual medical diagnoses. Echocardiography provides a more comprehensive set of diagnostic information for CHD, but its accessibility is limited in regions with scarce healthcare resources due to its time-consuming nature, high cost, and the need for substantial technical proficiency~\cite{wang2021automated}. If it were possible to combine the convenience of auscultation with the comprehensive capabilities of echocardiography, it would enable the screening of a broader population and provide physicians with more extensive reference information.

However, existing HSD datasets are limited to a few diseases and are strictly categorized as multi-class classification problems, which do not well align with medical practice (multi-label classification). Besides, such datasets overlook echocardiography due to its difficult-to-obtain nature \cite{wang2021automated}, which provides a more comprehensive set of diagnostic information yet. Hence, we aim to combine the convenience of auscultation with the comprehensive capabilities of echocardiography, enabling the screening of a broader population and providing physicians with more extensive reference information.

In this paper, we propose a new benchmark named HSDreport, which includes 2,~275 segments of heart sounds and their corresponding echocardiography. Each heart sound segment spans approximately 75 seconds across five body sites. Since our dataset is derived from practical applications, it presents unique two challenges compared to previous HSD datasets. Firstly, as echocardiography is composed by physicians based on images from ultrasound scans instead of directly derived from heart sounds, it poses a challenge to discard information that is difficult to discern from the heart sounds alone. Secondly, due to the natural language form of the report and the varying writing styles of different physicians, noise is introduced into the report. To address these issues, we leverage the strong semantic understanding capabilities and extensive internal knowledge of large language models (LLMs) \cite{achiam2023gpt} to extract abnormalities, ultimately constructing a 12-category multi-label benchmark.

To utilize the unique characteristics in our HSDreport, we have found that existing heart sound models \cite{chen2023robust,cheng2023heart,guo2023ds} fail to tackle the multi-label classification problem. This is because these models are primarily designed for multi-class classification and are limited to a few fixed classes.
Hence, we innovatively introduce a knowledge-aware query-based transformer for HSDreport to align the audio and text modalities~\cite{kritharoula2023large,kim2022joint,xiao2022rethinking,zhang2023knowledge} and leverage the extensive knowledge from medical pre-trained models by feeding abnormalities in textual expression. Moreover, we replace the typical phrase-based category input with detailed medical descriptions, enabling the model to acquire a more comprehensive understanding of medical diseases, their symptoms, and their characteristics, thereby facilitating a holistic view of the disease and enhancing classification accuracy. Considering the diversity in medical expressions, we further employ LLMs to provide multidimensional and varied descriptions of the same disease, thus aligning our approach more closely with medical practice and enhancing its robustness.

Our contributions are summarized as follows:
\begin{itemize}
%\item 我们定义了一个新的任务，即利用心音预测超声报告的信息，希望能够综合听诊的便利性和超声报告的全面性，并为此收集了一个全新的数据集。

%\item 我们提出了一种knowledge-aware的query-based transformer方法来应对新任务的挑战，并创新地使用疾病描述作为输入，使得模型对于疾病有着更准确的了解，另外在inference阶段综合利用LLM生成多维度多样化的描述的结果，进一步提升模型鲁棒性。

%\item Comparative analysis with 先进的心音模型 on 我们的 comprehensive dataset of real-world 心超 reports demonstrates 我们方法的 superior performance, showcasing notable improvements in all metrics。

\item \textbf{A benchmark for heart sound diagnosis.} 
We propose HSDreport, a practical and challenging benchmark for heart sound diagnosis. HSDreport combines the auscultation with echocardiographic analysis for a more accurate heart sound diagnosis, aiming to combine the convenience of auscultation with the comprehensiveness of echocardiographic analysis, which is not present in previous datasets and is highly significant for medical practice.

\item \textbf{A knowledge-aware approach for HSD.} We propose a knowledge-aware query-based transformer for HSD task. In addition, we use abnormality descriptions as inputs to enhance the model's accuracy in abnormality understanding. Furthermore, during inference, we comprehensively utilize the multidimensional diverse descriptions generated by the LLM, thereby further enhancing model robustness.

 %\textcolor{red}{We propose a knowledge-aware query-based transformer for HSD task. We leverage the extensive knowledge from medical pre-trained models and the internal knowledge of large language models to empower the audio-text alignment in HSD. This significantly improves the accuracy and robustness of classification. }

\item \textbf{Comprehensive performance evaluation.} We perform comparative analysis with state-of-the-art heart sound models on our comprehensive benchmark, showcasing notable improvements across all metrics. This study is the first to demonstrate the feasibility of using heart sounds to infer information from echocardiography, also proving our model's ability to effectively utilize knowledge to address multiple challenges.
\end{itemize}

%Cardiac auscultation remains a vital component of clinical medicine and has been widely used to screen for CHD due to its cost-effective. However, it heavily relies on clinical expertise and the acoustic hearing range of human ear~\cite{montinari2018history,mangione1997cardiac}. Echocardiography has demonstrated enhanced diagnostic efficacy over auscultation in CHD, while its accessibility in regions with limited healthcare resources is constrained by its time-consuming, high cost and stringent technical proficiency~\cite{wang2021automated}. Thus, if we can develop automated methods to achieve the diagnostic efficacy of echocardiography via auscultation, it will improve the early detection of disease and enable the screening of a wider population than possible with manual screening.
\section{Related Works}
\subsection{Heart Sound Diagnosis}
%当前心音数据集主要针对二分类以及少量类别的多分类问题。\cite{clifford2016classification}针对诊断结果，将音频划分为正常，异常，uncertain三类。\cite{oliveira2021circor}从心杂音的角度入手，将心音划分为了是否存在杂音的二分类问题，同时给出了杂音的部位，timing等信息。\cite{yaseen2018classification}收集了互联网上存在心音音频与对应标注的数据，构建了一个五分类数据集。针对这些数据集，当前的方法主要流程为音频特征提取，主体网络，分类网络。\cite{chen2023robust}首先对音频进行了去噪处理，然后通过STFT得到音频频谱图，他的主体网络基于CNN构建，并在末尾添加了attention模块。\cite{guo2023ds}组合使用了high-order spectral estimation和STFT作为音频特征，并设计了一种dual-stream convolutional neural network作为主体网络。\cite{cheng2023heart}组合使用了可训练的1d conv来提取特征并基于transformer构建了其主体网络。

The current datasets on heart sounds primarily target binary classification and multi-class classification for a limited number of categories. \cite{clifford2016classification} categorized heart sound recordings into three classes based on diagnostic outcomes: normal, abnormal, and uncertain. \cite{oliveira2021circor} approached heart sounds from the perspective of murmurs, classifying them into a binary category of presence or absence of murmurs, and additionally provided information on the location and timing of these murmurs. \cite{yaseen2018classification} collected heart sound recordings and corresponding annotations available on the internet to construct a five-category dataset. Regarding these datasets, the prevailing methodologies primarily consist of audio feature extraction, a main network, and a classification network. \cite{chen2023robust} initially applied noise reduction to the audio, then used Short-Time Fourier Transform (STFT) to obtain the spectrogram of the audio. Their main network was built on a CNN and incorporated an attention module at the end. \cite{guo2023ds} employed a combination of high-order spectral estimation and STFT for audio feature extraction, and designed a dual-stream CNN as the main network. \cite{cheng2023heart} simplified pre-processing, automatically extracting features using a one-dimensional convolution and built their main network based on a transformer architecture \cite{vaswani2017attention}. In contrast to previous datasets, our dataset initiates with echocardiography reports, which serve as the gold standard for HSD, containing a wealth of information essential for medical practitioners. In this paper, we process these reports into a 12-category multi-label task. Methodologically, we depart from the traditional paradigms of heart sound classification models by constructing a knowledge-aware, query-based transformer, which significantly enhances the model's performance.

%不同于以往的数据集，我们的数据集从心超报告入手，因为心超报告为HSD的金标准，拥有丰富的医生需要的信息，在本文中我们将心超报告处理为了12类multilabel任务。在方法上来说，我们打破以往心音分类模型范式，构建了一个knowledge-aware的query-based transformer，有效提升了模型性能。

%query-based transformer 

\subsection{Query-based Transformer}
Query-based transformers \cite{ma2023non} \cite{dan2021effects} \cite{lopez2023combining}
use adjustable query embeddings to make predictions and benefit from global attention, enabling them to gather information from an entire input. This allows them to outperform convolutional networks in terms of results. \cite{carion2020end} first introduced the query-based transformer in the object detection task and viewed it as a direct set prediction problem. \cite{li2022dn} introduces the concept of incorporating noised ground-truth boxes as positional queries in denoising training, an approach that has been shown to accelerate detection speeds. In addition to detection targets, \cite{cheng2022masked} employs mask attention for segmentation by utilizing predicted masks as attention masks, which enhances query refinement more efficiently than other query-based models. In the HSD task, we are the first to introduce the query-based transformer as the principal architecture. Unlike the models in other tasks, our queries consist of medical descriptions rather than words. Additionally, during the inference stage, we proposed a method that comprehensively utilizes multi-dimensional and diversified descriptions.

%在HSD任务中我们首次引入了query-based transformer作为主体结构，并与其他任务中的模型不同，我们的query为医学描述而非词语，且在inference阶段我们提出了一种综合利用多维度，多样化描述的方法。

\section{HSDreport: New Benchmark for Heart Sound Diagnosis}

%背景，任务定义。输入维度等等
\subsection{Background}
%心音作为廉价的先心病检测方式广为使用，然而当前HSD数据集只针对正常异常的情况，在医疗实践中能够提供的信息有限。心超能够提供更全面的信息，但是获取难度成本较大。因为心音与心超的相关性，并且为了结合心音的廉价获取性与心超的全面性，我们提出HSDreport，希望利用配对的心音与心超报告，从心音中学习心超报告中的关键信息。

The use of heart sounds as a cost-effective method for detecting congenital heart disease (CHD) is widely recognized. However, the current HSD dataset is limited to distinguishing between normal and abnormal conditions, which provides limited information in clinical practice. Echocardiography, on the other hand, offers more comprehensive information but is costly and difficult to obtain. Given the correlation between heart sounds and echocardiography, and to leverage the affordability of heart sound acquisition along with the comprehensiveness of echocardiography, we propose the HSDreport. This initiative aims to utilize paired heart sounds and echocardiography reports to extract key information from echocardiography reports based on heart sounds.

%具体的说，本任务要求输入一段心音a(维度为x*1)，和对应超声报告中提取的关键abnormalties D1到Di，分别用心音a去预测D1到Di

Given an input heart sound vector $\bf{h} \in \mathbb{R}{^{t}}$, we aim to predict the key abnormalities $E = \{ {E_1},{E_2},...,{E_k}\} $ extracted from the corresponding echocardiography report. The relationship can be modeled as:
%\begin{equation}
%\begin{aligned}
%\hat{D}_1 &= f_1(h), \\
%\hat{D}_2 &= f_2(h), \\
%&\vdots \\
%\hat{D}_k &= f_k(h),
%\nonumber
%\end{aligned}
%\end{equation}

\begin{equation}
{\hat{E}_i} = {f_i}\left( \bf{h} \right),i \in \left[ {1,k} \right]
\nonumber
\end{equation}
Here \( \hat{E}_i \) represents the predicted value of abnormality \( E_i \), and \( f_i \) is a predictive function for each \( E_i \) that maps from the heart sound input \( \bf{h} \) to the output space of abnormalities.

\begin{figure*}
    \centering
    \includegraphics[width=\linewidth]{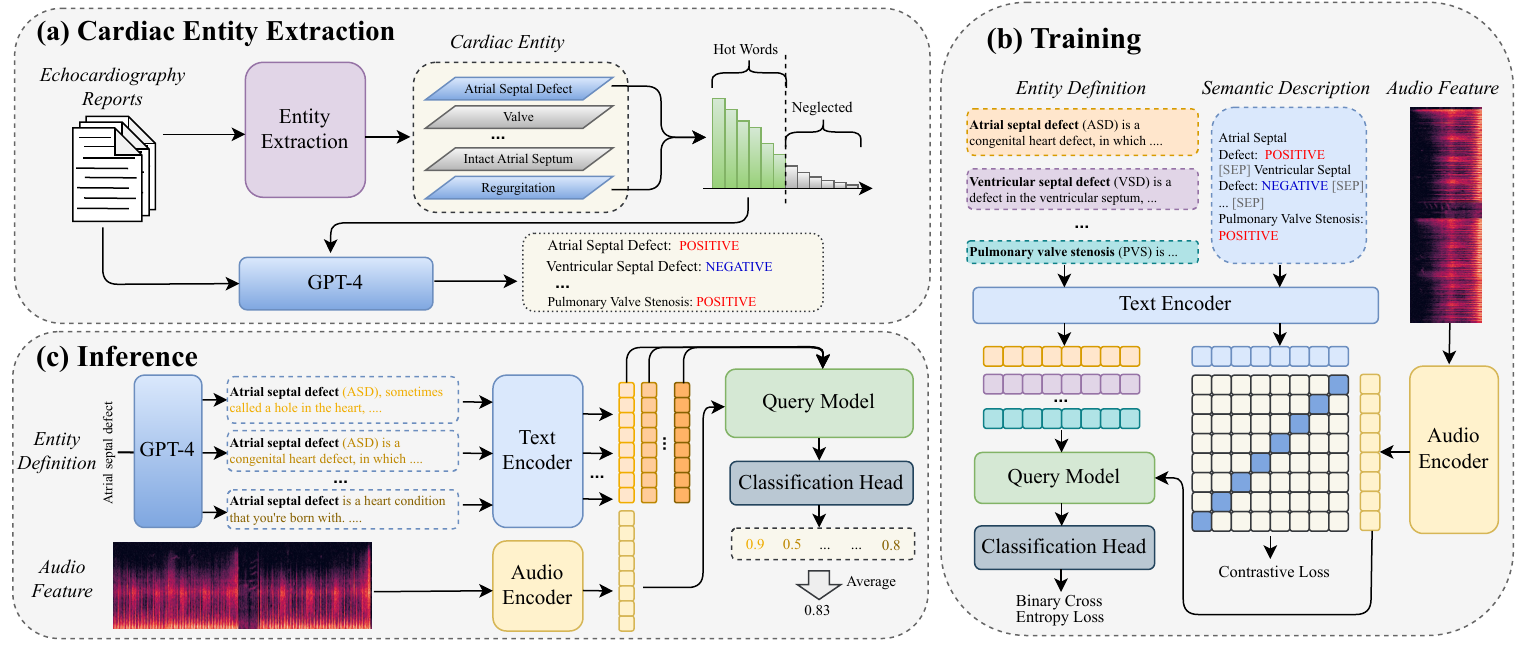}
    \caption{Framework overview of HSDreport, which consists of \textbf{(a) Cardiac Entity Extraction} to filter the hot words from the abnormal cardiac entities and verify the existence of them with GPT-4, \textbf{(b) Training} stage to train our model with extracted entity definitions and semantic descriptions paired by the heart sounds, and \textbf{(c) Inference} stage to obtain the diagnosis for a specific entity with various definitions derived from GPT-4.}
    % \label{fig:enter-label}
\end{figure*}

%data collection
%data processing
\subsection{Data Collection}
%To bridge the gap between existing binary classification heart sound datasets and medical practice, we propose a task using heart sound to directly predict echocardiography reports. For this purpose, we collect a dataset, AuscultCardiography, 

HSDreport includes 2, 275 participants and 2, 275 auscultation recordings. Specifically, digital auscultation recordings of heart sounds from patients aged $\leq$18 years who had undergone echocardiography, are collected.
% in a multicenter retrospective study.  
The auscultation protocol consists of recordings over 5 body sites: aortic region (right 2nd intercostal space), pulmonic region (left 2nd intercostal space, parasternal), Erb's point (left 3rd intercostal space aka left lower sternal border), tricuspid region (left 4th intercostal space, parasternal), mitral region (left 5th intercostal space, midclavicular). To detect sufficient cardiac cycles, at least 15s of heart sound using direct skin contact is obtained per site. The electronically amplified stethoscope (Littmann 3200, 3M) is used for data acquisition. During the examination the participant is seated, laid down, or held to the most comfortable position. A complete transthoracic echocardiography is available for all subjects using standard views and techniques according to established guidelines. The report is written by professional ultrasound technicians.

%Data quality assessment is performed.首先，所有investigator相信不应该被包括的病人的数据都已被去除，这包括心音极难听到的低质量数据，或者不适合加入的数据（如患者年龄大于18岁或没有echocardiography的情况）。下一步investigator对echocardiography进行了检查，all entries are screened for incorrectly entered or measured values, inconsistent data or outliers, and deleted as appropriate. 虽然我们执行了以上步骤来确保data quality，但需要注意的是 the sounds are recorded in an ambulatory environment. 所以 A variety of noisy sources是必然存在的, including stethoscope rubbing noise, speaking, crying, or laughing sounds in the background. On the other hand, the proposed dataset is a representative sample of actual settings in which machine-based auscultation systems may function automatically.

During the data collection process, investigators have taken the following steps to improve the quality of the data collected. Firstly, all investigators remove data they deem inappropriate, including low-quality data where heart sounds are difficult to hear, data without corresponding echocardiography, and data from patients over 18 years old. In the next step, investigators review the echocardiography entries, screening all entries for incorrectly entered or measured values, inconsistent data, or outliers, and delete such entries as appropriate. Despite these steps taken to ensure data quality, it is important to note that the heart sounds are recorded in an ambulatory environment. Therefore, a variety of noisy sources are inevitably present, including stethoscope rubbing noise, and background sounds such as speaking, crying, or laughing. From another perspective, however, this dataset is more closely aligned with medical practice.

\subsection{Data Processing}
%一份心超报告通常由三部分构成：数值指标，超声描述和超声诊断。心超报告是由医师阅读图像模态写得的报告，虽然其中医师测量得到的各种血流速度，厚度等数值指标仍与心音紧密相关，即不同速度，厚度等情况下心音可能会有区别，但在当前数据量的情况下可能难以直接用心音测量具体的数值指标。因此我们首先从报告中去除了所有数值信息。我们认为这么做并不会损失关键信息，因为医师对于异常的指标，会以自然语言的形式在超声描述或诊断中进行叙述。
A typical echocardiography report is composed of three primary sections: numerical indices, description, and diagnosis. Echocardiography reports present challenges for direct learning by models, thus necessitating an approach where we extract all abnormalities from an echocardiography as task annotations through a series of steps, detailed further below and in Figure 1(a). Firstly, this report is produced by physicians who interpret various imaging modalities. Although various numerical indices such as blood flow velocities and thicknesses are closely related to heart sounds—where different velocities and thicknesses may correspond to variations in heart sounds—it is challenging to directly measure specific numerical indices solely through heart sounds given the current data size. Therefore, we have initially excluded all numerical information from the report. We believe this approach does not result in the loss of critical information, as physicians tend to describe any abnormal indices using natural language within the description or diagnosis sections.

%除此之外，由于心超报告的描述和诊断部分由医师使用自然语言写成，且没有固定的模板，因此心超报告存在很大的噪声，典型的特征是超声描述中大部分内容均描述的是正常的部分，abnormalties很稀疏，因此难以让模型直接学习超声描述部分。而在诊断部分只有最终的疾病诊断结果，而无与其相关的其他abnormalties，存在信息不完整的情况，因此也不适合让模型直接进行学习。因此我们认为需要结合超声描述与超声诊断部分，并去噪得到所有关键abnormalties信息供模型学习。为此我们首先使用了医学entity extraction方法来提取所有报告里的医学entity，并将这些词出现的数量进行排序，然后我们去除了这些entity中非abnormalties的词，如部位和描述正常情况的词，然后选取了12类频率出现最高的abnormalties来代表一份超声报告可能存在的异常信息，剩下的abnormalties由于出现数量低于20次，难以由模型学习，因此简便起见in this paper我们省略了他们。

Secondly, since the description and diagnosis sections of echocardiography reports are written by physicians using natural language without a fixed template, these reports contain significant noise. A typical characteristic is that most of the content in the description sections pertains to normal findings, with abnormalities being quite sparse. Consequently, it is challenging for models to directly learn from the description sections. Moreover, the diagnosis sections only contain the final disease diagnoses without the associated abnormalities, 
resulting in incomplete information. Therefore, we believe it is necessary to integrate both the description and diagnosis sections and denoise them to extract all key abnormalities for model learning. To achieve this, we first employ a medical entity extraction method to extract all medical entities from the reports and rank these entities by their frequency of occurrence. Subsequently, we remove non-abnormality entities and select the 12 most frequently occurring abnormalities to represent the potential abnormal information in an echocardiography report. The remaining abnormalities, which appear fewer than 20 times, are excluded from model training due to their infrequency. 

%下一步我们使用了LLM去寻找了每份报告中是否存在这12类abnormalties，需要进行这一步的原因依然由于超声描述和超声诊断的自然语言性质，因此很多abnormalties在其中存在不规范的形式，因此需要利用LLM的internal knowledge和自然语言理解能力来准确寻找12类abnormalties存在与否，对于医师没有提及的abnormalies，我们均认为其不存在。

The final step involves utilizing a LLM to identify the presence of 12 categories of abnormalities in each report. This step is necessary due to the inherent natural language characteristics of description and diagnosis, which often result in irregular forms of abnormalities. Consequently, the internal knowledge and natural language understanding capabilities of the LLM are leveraged to accurately detect the existence of these 12 types of abnormalities. For any abnormalities not mentioned by the physician, we assume their absence.

\section{Methodology}
%现有心音相关方法主要针对二分类任务，因此不能很好地应对新任务带来的报告噪声多，医学术语多的挑战，为此我们设计了全新的处理流程来处理这一任务，如图1所示，新方法由三个部分组成，包括LLM赋能的Cardiac entity extraction步骤，结合医学entity definition以及医学预训练模型的query-based transformer进行的training步骤，和LLM赋能以综合使用多维度，多样化医学entity definition进行的inference步骤。 In the following sections, we will describe the them in detail.

%Existing heart sound-related methods primarily address binary classification tasks. Consequently, they struggle to effectively manage the challenges posed by the new task,这主要是本数据集的复杂程度带来的. To tackle this issue, we have designed an innovative processing workflow, as illustrated in Figure 1(b-c). The new method consists of two components: (1) A training step that integrates medical entity definitions, medical data pre-trained text encoder and a query-based transformer model. (2) An inference step powered by LLMs to comprehensively utilize multi-dimensional and diverse medical entity definitions. In the following sections, we will describe these components in detail.

\subsection{Training}
%，我没有使用传统的心音相关模型中的特征提取-主网络-线性层分类的结构，而是创新地基于query-based transformer构建了我们的模型，从而方便地引入医疗知识预训练text encoder来编码每个abnormal entity。我们认为这么做将会更好地对齐两个模态。与以往query-based transformer使用词语作为类别不同，由于医学短语在实际中往往存在生僻和表达多样的问题，我们创新性地使用医学描述代替词语作为类别，这可以使得模型对于医学疾病更详细的了解其症状和特点，从而有一个holistic view of the disease。因此对于每个abnormal entity${e_i} \in \mathbb{R}{^{x \times 1}}$会有: t=f_text(f_define(ei))属于。

In the development of the HSDreport, we diverge from the conventional architectural paradigm prevalent in previous heart sound models, which typically entails feature extraction
% , a principal network, 
and a linear classification layer. Instead, we innovatively constructed our model based on a query-based transformer. This approach facilitates the integration of a medically pre-trained text encoder to encode each abnormal entity. We argue that this method will enhance alignment between the two modalities. Contrary to the traditional use of query-based transformers, where words are employed as categories, we employ medical descriptions rather than words due to the frequent obscurity and variability in the expression of medical phrases in practice. This innovative substitution allows the model to gain a more detailed understanding of the symptoms and characteristics of medical conditions, thereby offering a holistic view of the disease. So for each abnormal entity ${E_j}$, the following applies:

\begin{equation}
{{\bf{t}}_{j}} = {f_{\rm{text}}}\left( {{f_{\rm{definition}}}\left( {{E_j}} \right)} \right) \in {\mathbb{R}^{ {d}}}
\nonumber
\end{equation}

%对于心音，我们使用FBank提取其频谱图，然后通过预训练的audio encoder得到其特征：。
Here $j$ represents the $j^{th}$ class. For the $i^{th}$ sample's heart sound ${{\bf h}_i} \in \mathbb{R}{^{t \times 1}}$, we first extract their spectrograms, then obtain their features through a pre-trained audio encoder:
\begin{equation}
{{\bf{H}}_i} = {f_{\rm{audio}}}\left( {{f_{\rm{feature}}}\left( {{{\bf{h}}_i}} \right)} \right) \in {\mathbb{R}^{{x_h} \times d}}
\nonumber
\end{equation}
%然后我们将具有语义的医学描述特征T_j作为query，和心音特征一起H_i送入query-based transformer，在通过由线性层组成的classification head后得到预测值${\hat{E}_{i,j}}$:
Then we employ the semantically enriched text features ${{\bf{t}}_{j}}$
as the query, which, together with the heart sound features ${{\bf{H}}_i}$, are fed into the query-based transformer. After processing through a classification head composed of linear layers, the predicted values ${\hat{E}_{i,j}}$ are obtained:
\begin{equation}
{\hat{E}_{i,j}} = {f_{\rm{linear}}}\left( {{f_{\rm{transformer}}}\left( {{\bf{H}}_i,{\bf{t}}_j} \right)} \right) \in {\mathbb{R}}
\nonumber
\end{equation}
Then we can compute the binary cross-entropy loss (BCE) as follows:

\begin{equation}
\small
\resizebox{\linewidth}{!}{
$
{{\mathcal{L}}_{BCE}} = 
- \frac{1}{{NK}}\sum\limits_{i = 1}^N {\sum\limits_{j = 1}^K {\left[ {{s_{i,j}}\log \left( {{{\hat E}_{i,j}}} \right) + \left( {1 - {s_{i,j}}} \right)\log \left( {1 - {{\hat E}_{i,j}}} \right)} \right]} }
$
}
\nonumber
\end{equation}
%为了进一步对齐音频与文本模态，我们决定使用样本的semantic description ri，它由文本形式的entity名称，标注以及分隔符[SEP]构成。在这里没有使用医学描述而使用名称的原因是将所有医学描述拼接起来长度过长。在ri通过text encoder后得到Ri，我们在它与音频特征H之间计算contrastive loss
Here ${s_{i,j}}$ is the label. To further align the audio and textual modalities, we choose to employ the semantic description \( R_i \), composed of the entity names in textual form, annotations, and the delimiter [SEP]. The rationale for using names instead of medical descriptions is that concatenating all medical descriptions results in excessive length. After \( R_i \) is processed through the text encoder to obtain ${{\bf{r}}_i} \in {\mathbb{R}^d}$, we compute the contrastive loss between ${{\bf{r}}_i}$ and the audio features ${\bf{H}}_i$:
\begin{equation}
\small
\resizebox{\linewidth}{!}{
$
{\mathcal{L}_{Con}} =  - \frac{1}{N}\sum\limits_{i = 1}^N {\left( {\log \frac{{{e^{\left\langle {{{\bf{H}}_i},{{\bf{r}}_i}} \right\rangle/\tau }}}}{{\sum\nolimits_{k = 1}^N {{e^{ \left\langle {{{\bf{H}}_i},{{\bf{r}}_k}} \right\rangle/\tau }}} }} + \log \frac{{{e^{ \left\langle {{{\bf{r}}_i},{{\bf{H}}_i}} \right\rangle/\tau }}}}{{\sum\nolimits_{k = 1}^N {{e^{ \left\langle {{{\bf{r}}_i},{{\bf{H}}_k}} \right\rangle/\tau }}} }}} \right\rangle} $}
\label{contrastive}
\end{equation}
%在这里\[\left\langle . , . \right\rangle \]会首先对Hi进行平均池化，然后计算cos相似度。tao为温度参数。

Here the operation \(\left\langle . , . \right\rangle\) first applies average pooling to \( \bf{H}_i \), followed by the computation of the cosine similarity and  \( \tau \) represents the temperature.
Finally, we sum the two losses to obtain the final loss, where the weight \(\lambda\) is a learnable parameter:

\begin{equation}
\mathcal{L} = {\mathcal{L}_{BCE}} + \lambda {\mathcal{L}_{Con}}
\label{loss}
\end{equation}

\subsection{Inference}
%考虑到医学表达多样化的特点，我们使用了大语言模型对同种疾病进行多维度，多样化的描述，并在inference阶段综合使用这些描述，从而使得我们的方法可以更接近医学实践并且鲁棒性更好。具体来说对于第j类的abnormal entity E_j，我们使用了LLM生成了p次描述，并将其一起输入模型：

Considering the diverse nature of medical expressions, we utilize an LLM to generate multi-dimensional and diverse descriptions of the same disease. During the inference stage, we integrate these descriptions to align our approach more closely with medical practice and enhance its robustness. Specifically, for the $j^{th}$ class of abnormal entity \(E_j\), we generate \(p\) descriptions using the LLM and input them collectively into the model:

\begin{equation}
{{\hat E}_{i,j}} = {f_{\rm{linear}}}\left( {{f_{\rm{t}}}\left( {{{\bf{H}}_i},{{\bf{t}}_{j,1}},...,{{\bf{t}}_{j,p}}} \right)} \right) \in {\mathbb{R}^p}
\nonumber
\end{equation}

%这里Hi为第i条样本的音频特征，tjp为针对第j类生成的第p个描述的特征，ft为query-based transformer.随后我们将Eij求平均得到该类的最终结果

Here \( {{\bf{H}}_i} \) represents the audio feature of the \( i^{th} \) sample, while \( {\bf t}_{j,p} \) denotes the feature of the \( p^{th} \) description generated for class \( j \). The function \( f_t \) refers to the query-based transformer. Subsequently, we compute the average of \( {{\hat E}_{i,j}} \) to obtain the final result for that class.

\begin{figure}[t]
    \centering
    \includegraphics[width=\linewidth]{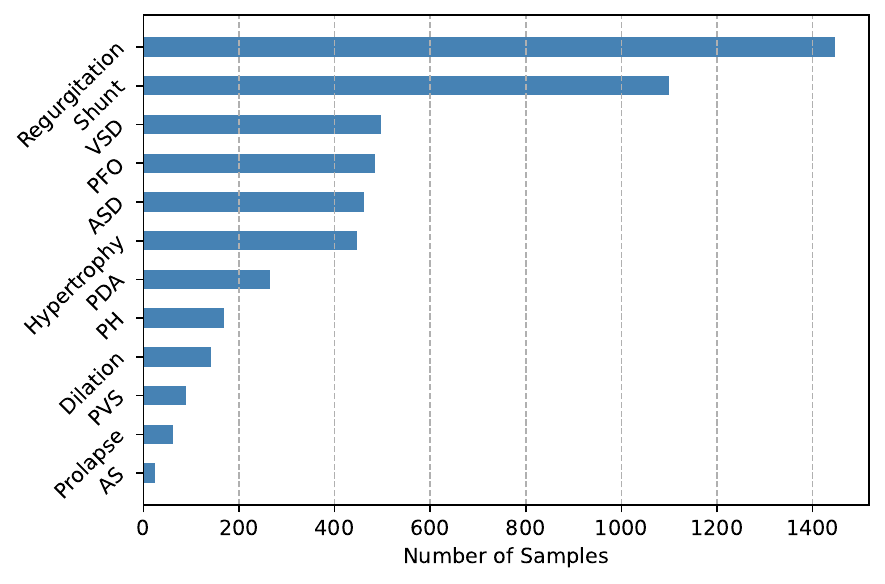}
    \caption{Data distribution. The total number of samples is 2275, and the numbers in the figure represent the number of positives in that category.}
    \label{fig:distribution}
\end{figure}

\begin{table*}[t]
    \centering
    \resizebox{\linewidth}{!}{
    \begin{tabular}
    {l|cc c c c c c c c c c c | c}
        \toprule
        Method & ASD & VSD & PVS & PDA & PFO & AS & PH & Prolapse & Regurgitation & Shunt & Hypertrophy & Dilation & Average \\
        \midrule
        \multicolumn{14}{c}{\textit{Precision}}\\
        \midrule
        GPT-4o & 38.04 & 37.83 & 48.83 & 44.57 & 41.09 & 49.35 & 46.30 & 48.26 & 18.48 & 24.35 & 40.87 & 46.96 & 40.33 \\
        STFT-HSC & \textbf{88.54} & 88.61 & 47.83 & \textbf{94.96} & \textbf{89.47} & 49.35 & 46.29 & 48.26 & 68.99 & 67.93 & 77.60 & 46.96 & 67.89  \\
        DS-CNN & 62.28 & 82.17 & 47.81 & 85.33 & 84.56 & 49.35 & 46.30 & 48.26 & 67.13 & 70.89 & \textbf{81.93} & 46.94 & 64.41 \\
        CTENN & 61.43 & 82.83 & 69.69 & 72.93 & 86.82 & \textbf{99.78} & \textbf{80.03} & 48.26 & 57.46 & 69.19 & 69.27 & \textbf{57.11} & 71.23  \\
        \midrule
        \textbf{Ours} & 68.78 & \textbf{90.28} & \textbf{84.31} & 93.65 & 78.48 & \textbf{99.78} & 76.89 & \textbf{86.39} & \textbf{73.40} & \textbf{72.35} & 81.62 & 51.57 & \textbf{79.79} \\
        \midrule
        \midrule
        \multicolumn{14}{c}{\textit{Recall}}\\
        \midrule
        GPT-4o & 50.00 & 50.00 & 50.00 & 50.00 & 50.00 & 50.00 & 50.00 & 50.00 & 50.00 & 50.00 & 50.00 & 50.00 & 50.00 \\
        STFT-HSC & 52.73 & 80.71 & 50.00 & 54.00 & 64.37 & 50.00 & 49.77 & 50.00 & 69.98 & 67.23 & 66.00 & 50.00 & 58.73  \\
        DS-CNN & 57.87 & 75.09 & 49.54 & 57.76 & 61.67 & 50.00 & 50.00 & 50.00 & 64.91 & 69.48 & 81.33 & 49.77 & 59.79\\
        CTENN & 61.14 & 78.98 & 73.41 & 60.78 & 65.32 & \textbf{83.33} & 55.65 & 50.00 & 57.85 & 68.88 & 65.06 & \textbf{52.65} & 64.42 \\
        \midrule
        \textbf{Ours} & \textbf{68.31} & \textbf{86.66} & \textbf{84.31} & \textbf{75.75} & \textbf{70.78} & \textbf{83.33} & \textbf{58.35} & \textbf{68.52} & \textbf{73.28} & \textbf{71.93} & \textbf{88.06} & 51.25 & \textbf{73.38} \\
        \midrule
        \midrule
        \multicolumn{14}{c}{\textit{F1 Score}}\\
        \midrule
        GPT-4o & 43.21 & 43.07 & 48.89 & 47.13 & 45.11 & 49.67 & 48.08 & 49.12 & 26.98 & 32.75 & 44.98 & 48.44 & 43.95 \\
        STFT-HSC & 48.70 & 83.67 & 48.89 & 54.75 & 68.53 & 49.67 & 47.96 & 49.12 & 69.20 & 66.70 & 69.14 & 48.43 & 58.73 \\
        DS-CNN & 58.53 & 77.62 & 48.66 & 60.78 & 64.81 & 49.67 & 48.08 & 49.11 & 65.38 & 68.70 & 81.62 & 48.31 & 60.11 \\
        CTENN & 61.28 & 80.61 & 71.36 & 63.84 & 69.50 & \textbf{89.89} & 58.18 & 49.12 & 57.43 & 68.61 & 66.62 & \textbf{53.35} & 65.82 \\
        %PANNs&56.51&70.49&58.08&61.88&70.43&74.78&63.40&59.11&68.15&69.11&72.77&54.67&64.94\\

        %BEATs&66.08&62.89&64.37&65.06&68.02&74.78&58.18&59.11&70.66&69.98&82.52&54.67&66.36\\

        %wav2vec2.0&56.88&61.02&58.08&61.17&65.97&69.67&59.81&48.89&68.69&65.36&76.98&54.67&62.26\\
        \midrule
        \textbf{Ours} & \textbf{68.53} & \textbf{88.27} & \textbf{84.31} & \textbf{81.78} & \textbf{73.57} & \textbf{89.89} & \textbf{61.80} & \textbf{74.33} & \textbf{73.34} & \textbf{71.64} & \textbf{84.20} & 51.35 & \textbf{75.25} \\
         \bottomrule
    \end{tabular}
    }
    \caption{The precision, recall, and F1 scores of GPT-4o, three state-of the-art models for HSD, and our method on the multi-label benchmark. Our model achieves significant improvements in all three metrics.}
    \label{tab:main}
\end{table*}

\section{Experiments}
\subsection{Experimental Setup}
\paragraph{Datasets}
The dataset utilized in our study comprises 2, 275 heart sounds paired with the echocardiography reports, each approximately 75 seconds long, split into training and test sets respectively in a 9:1 ratio. The extracted abnormal entities from the paired reports are filtered and categorized into 7 diseases: Atrial Septal Defect (ASD), Ventricular Septal Defect (VSD), Pulmonary Valve Stenosis (PVS), Patent Ductus Arteriosus (PDA), Patent Foramen Ovale (PFO), Aortic Stenosis (AS), Pulmonary Hypertension (PH), and 5 symptoms: Prolapse, Regurgitation, Shunt, Hypertrophy, and Dilation. The distribution is demonstrated in Figure~\ref{fig:distribution}.

\paragraph{Baselines}
We adopt three state-of-the-art heart sound auscultation methods as our baselines: STFT-HSC~\cite{chen2023robust}, DS-CNN~\cite{guo2023ds} (both CNN-based), and CTENN~\cite{cheng2023heart} (transformer-based). To enable them to adapt to our multi-label classification task, we have modified their classification heads and retrained them following the instructions. Due to the existing LLMs' capability for audio processing, we also adopt GPT-4o\footnote{\url{https://openai.com/index/hello-gpt-4o/}} as an LLM baseline.

\paragraph{Training Stage}
At the training stage, we utilize the filter banks of heart sounds as the audio features, with a frame length of 100ms and a frame shift of 40ms. Chunk dropping, speed perturbation, clipping, noising, amplifying, and SpecAugment~\cite{park2019specaugment} are applied with the SpeechBrain library as data augmentation. The definition for each entity is derived from Wikipedia. For the text encoder and audio encoder, we adopt pre-trained PubMedBERT~\footnote{\url{https://huggingface.co/NeuML/pubmedbert-base-embeddings}} and ResNet50~\footnote{torchvision.models.resnet50} as the initialization. We use an AdamW optimizer with a learning rate of 5e-5 and a weight decay of 0.02. We train on an A100 GPU for 100 epochs with 20 for warming up, and the cosine learning rate schedule is applied. $\lambda$ in Equation~(\ref{loss}) is initialized to 1.

\paragraph{Inference Stage}
At the inference stage, we utilize GPT-4 to generate 100 descriptions for each class. We adopt precision, recall, and F1 scores as the evaluation metrics and set 0.5 as the discriminant threshold.

\subsection{Results}
\subsubsection{Multi-label Classification}
We conduct experiments on the proposed multi-label benchmark HSDreport and compare the diagnosis performance with the state-of-the-art approaches. The precision, recall, and F1 scores of each category and the averaged values are reported. As demonstrated in Table~\ref{tab:main}, Our approach attains the highest F1-scores in nearly all categories and exceeds the best baseline by 9.4\%, demonstrating the effectiveness of the proposed knowledge-aware, query-based transformer for HSD task. Among the baselines, the transformer-based method (CTENN) outperforms those based on CNNs (STFT-HSC and DS-CNN). In addition, GPT-4o struggles with this task, achieving the lowest scores. We attribute this performance to the limited size of heart sound data in the public dataset and the absence of annotations similar to those in the HSDreport, which prevents GPT-4o from generalizing to our benchmark.

Ulteriorly, we split the evaluation metric into precision and recall, where recall is particularly important for medical diagnosis. Obviously, our approach exhibits the most superiority in the recall score for various diseases and symptoms. Such a superiority means that our method can reduce the false dismissal rate and avoid disease misdiagnosis effectively. It is also noticed that all of the methods are not able to recognize the existence of Dilation effectually, and we guess it's because the characteristics of Dilations may hardly be identified by heart sounds.

\begin{figure}[ht]
    \centering
    \includegraphics[width=\linewidth]{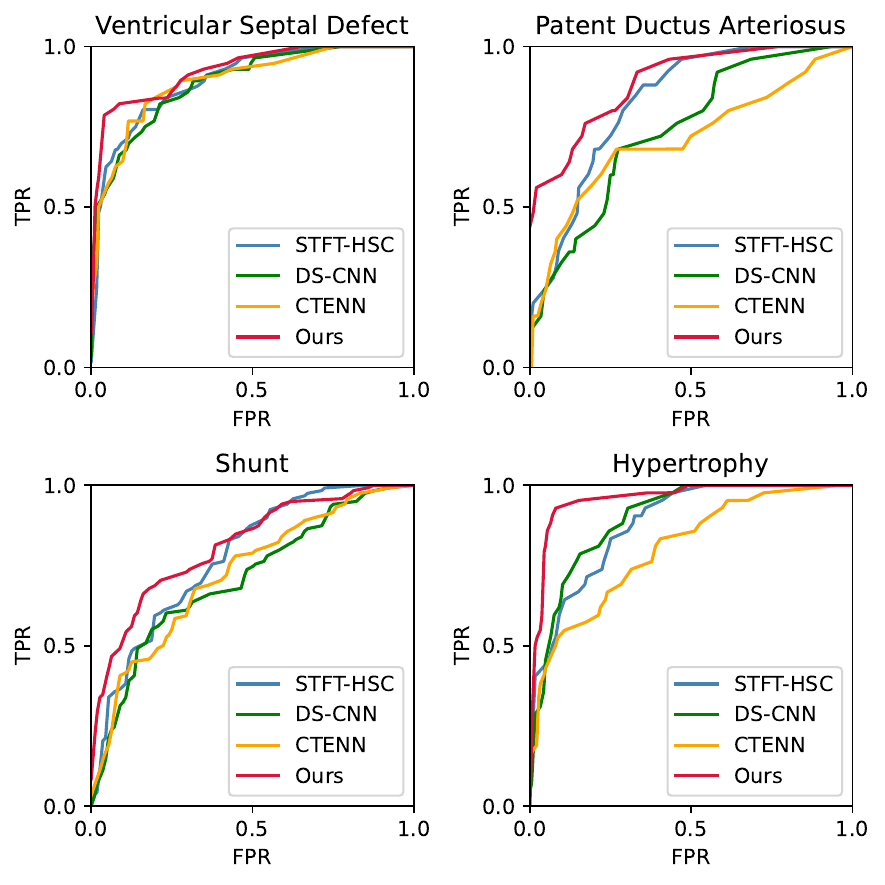}
    \caption{ROC curves of four classes in the benchmark: VSD, PDA, Shunt, and Hypertrophy.}
    \label{fig:roc}
\end{figure}

\subsubsection{ROC Curve}
To further analyze the robustness under different discrimination thresholds, we plot the receiver operating characteristic (ROC) curve of diseases (VSD and PDA) and symptoms (Shunt and Hypertrophy) respectively. As shown in Figure~\label{fig:roc}, the area under the ROC curve (AUC) for our approach is notably superior and maintains the highest values compared to that of the established baselines, which not only signifies the discriminatory capacity of our model between positive and negative cases but also underscores its robustness and generalizability across different threshold settings.

\begin{table*}[]
    \centering
    \resizebox{\linewidth}{!}{
    \begin{tabular}
    {l|cc c c c c c c c c c c | c}
        \toprule
        Method & ASD & VSD & PVS & PDA & PFO & AS & PH & Prolapse & Regurgitation & Shunt & Hypertrophy & Dilation & Average \\
        \midrule
        Ours & \textbf{68.53} & \textbf{88.27} & \textbf{84.31} & \textbf{81.78} & 73.57 & \textbf{89.89} & 61.80 & \textbf{74.33} & 73.34 & 71.64 & 84.20& 51.35 & \textbf{75.25} \\
        \midrule
        - Text Encoder & 68.50 & 84.46 & 83.69 & 79.03 & 72.22 & \textbf{89.89} & 61.09 & 67.40 & \textbf{76.48} & 72.09 & 86.41 & \textbf{52.95} & 74.52\\
        - Audio Encoder& 65.39& 82.56& 69.22& 68.96& 65.04& 49.56& \textbf{64.97} & 70.53& 66.23& 71.24& 82.87& 52.94& 67.46\\
        - $\mathcal{L}_{Con}$ & 66.78& 87.40& 67.56& 78.19& \textbf{75.23}& 83.11& 64.04& 67.40& 74.44& \textbf{74.30} & 85.94& 47.72& 72.68\\
        % - $\mathcal{L}_{focal}$ \\
        - Entity Definition &55.28& 82.29& 58.18& 74.82& 64.90& 48.77& 61.85& 63.40& 72.20& 73.46& \textbf{89.46} & 52.26& 66.41\\
         \bottomrule
    \end{tabular}
    }
    \caption{This table presents an ablation study to assess the validity of the text encoder, audio encoder, contrastive loss, and entity definition input during the training stage. F1 scores are reported for each category. The table demonstrates the efficacy of all modules.}

    \label{tab:ablation_train}
\end{table*}

\begin{table*}[]
    \centering
    \resizebox{\linewidth}{!}{
    \begin{tabular}
    {l|cc c c c c c c c c c c | c}
        \toprule
        Text Input & ASD & VSD & PVS & PDA & PFO & AS & PH & Prolapse & Regurgitation & Shunt & Hypertrophy & Dilation & Average \\
        \midrule
         Only Entity & \textbf{69.27} & 87.40 & 80.78 & 77.95 & 73.04 & 71.66 & 61.80 & 70.53 & 71.06 & 70.32 & 84.97 & 55.81 & 72.88 \\
         $N=1$ & 68.12 & 87.57 & 76.87 & 79.92 & 70.52 & \textbf{89.89} & 61.09 & 65.77 & \textbf{73.74} & 71.28 & \textbf{84.20} & \textbf{59.69} & 74.05\\
         $N=10$ & 68.53 & \textbf{88.27} & \textbf{84.31} & \textbf{81.78} & \textbf{73.57} & \textbf{89.89} & \textbf{61.80}& \textbf{74.33} & 73.34 & 71.64 & \textbf{84.20} & 51.35 & \textbf{75.25}   \\
         $N=50$ & 68.12 & 87.57 & \textbf{84.31} & 79.92 & 73.04 & \textbf{89.89} & \textbf{61.80} & 67.40 & 73.34 & \textbf{72.42} & \textbf{84.20} & 58.68 & 75.05\\
         \bottomrule
    \end{tabular}
    }
    \caption{This table presents an ablation study aggregating descriptions generated by different quantities of GPT-4 during the inference stage. The F1 scores are reported for each category. The table demonstrates that using ten descriptions is the most effective method. }
    \label{tab:ablation_inference}
\end{table*}

\subsection{Ablation Study}
We conduct ablation studies for the training and inference stage respectively to verify the effectiveness and robustness of each counterpart in the proposed method.

\subsubsection{Training Stage}
We first establish the ablation by systematically disabling components during the training stage. This involves, specifically, replacing the pre-trained initializations of both the text encoder and audio encoder with random initialization procedures individually. Additionally, we excise the contrastive loss component from the overarching loss function outlined in Equation~\ref{contrastive}. Lastly, we substitute the comprehensive entity definitions with the entity words in our model's architecture. The outcomes of these ablation experiments are documented in Table~\ref{tab:ablation_train}, revealing the indispensable nature of each module in the diagnostic process as any module removal brings a distinct decline in performance. Among these, replacing the entity definition exerts the most significant impact, substantiating our assertion that medical terminology is characterized by rare and diverse issues, with the lexical semantics being insufficient for models to comprehend diseases. Replacing the audio encoder also has a substantial effect, likely due to the complexity of audio features, which are difficult to learn from a random state given the current sample size. In contrast, replacing the text encoder and contrastive loss has a relatively minor impact.%在这些模块中间，替换entity definition有着最大的影响，这证明了我们的观点，即医学术语存在生僻多样的问题，词汇的语义不足以让模型理解疾病。替换audio encoder同样有着很大的影响，这可能是因为音频特征较为复杂，难以在当前样本量的情况下从随机状态学习而得。相比以上两个模块，替换text encoder和clip loss的影响相对较小。

%, which can be acutely observed for PVS and Prolapse. Notably, the initialization strategy for the audio encoder and the strategic shift from using the entity directly to incorporating its detailed definition emerge as pivotal factors, exerting a substantial influence on the model's performance across various diagnostic categories.

\begin{figure}
    \centering
    \includegraphics[width=\linewidth]{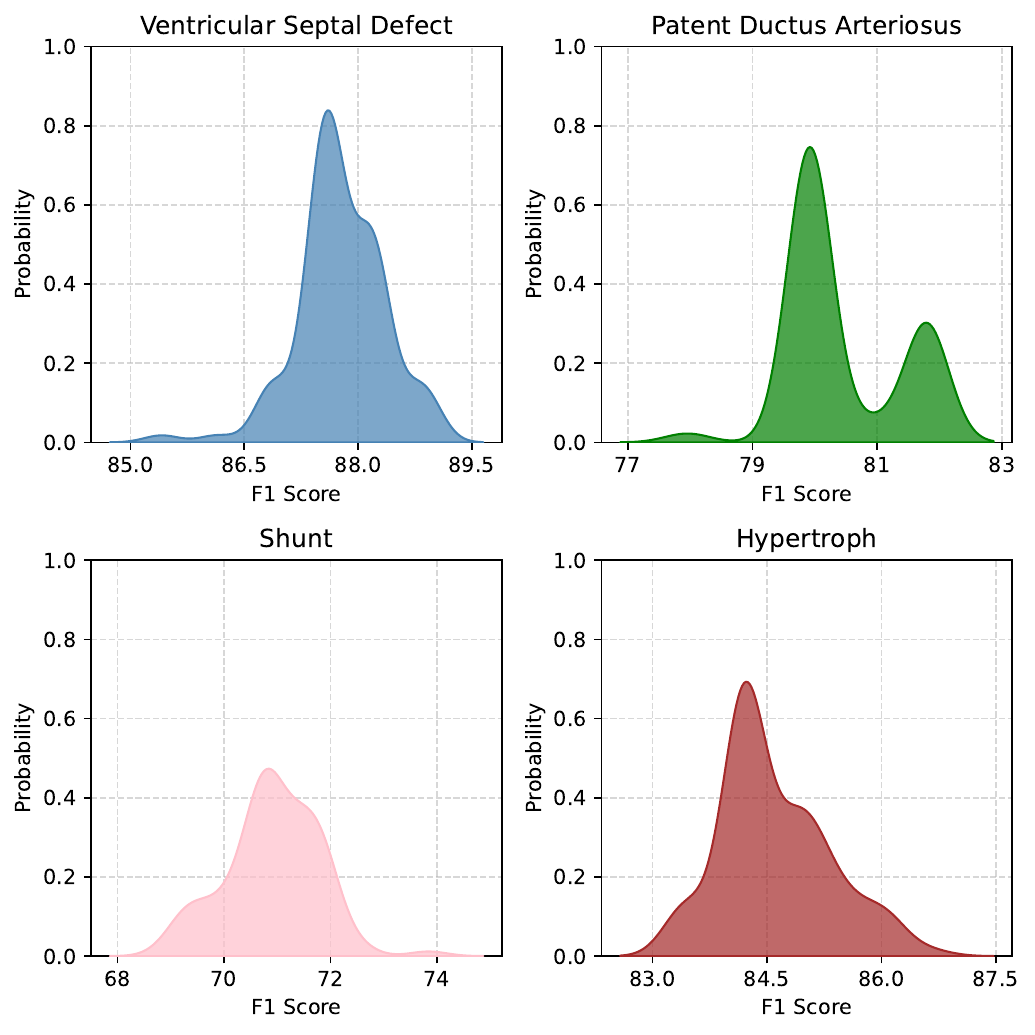}
    \caption{The distribution of F1 scores for VSD, PDA, Shunt, and Hypertrophy categories.}
    \label{fig:f1_distribution}
\end{figure}

\subsubsection{Inference Stage}

%As for the inference stage,我们验证了使用不同数量的gpt4生成的entity defination下的结果. In this regard, we replace the text input for the inference stage with the entity words and different numbers of generated entity definitions and record the performance as Table~\ref{tab:ablation_inference} shows. It is obvious that adopting the entity definition as text query consistently outperforms only utilizing the entity words.这同样证明了在HSD任务中医学描述具有更准确语义的这一特点。 Besides, such a performance enhancement can be further boosted by increasing the number of entity definitions, which stabilizes when $N=10$, and this is also the default setting in our experiments.在使用更多的entity definitions时我们没有关注到进一步的结果变化，我们认为这是由于10条描述对于大部分类别来说，gpt4已生成了足够多样，丰富的描述。

During the inference stage, we validate the outcomes under various quantities of GPT-4-generated entity definitions. Specifically, we substitute the text input for the inference stage with the entity words (which means $N=0$) and varying numbers of generated entity definitions and document the performance as shown in Table~\ref{tab:ablation_inference}. It is evident that using the entity definition as a text query consistently yields superior results compared to merely utilizing the entity words. This also substantiates the usefulness of medical descriptions in the HSD task, which possess more precise semantics. Additionally, such performance enhancement can be further amplified by increasing the number of entity definitions. This enhancement stabilizes when $N=10$, which also serves as the default setting in our experiments. No further changes in outcomes were observed with an increase in the number of entity definitions beyond this point, likely because ten descriptions are sufficiently diverse and rich for most categories as generated by GPT-4.

To further analyze the generality of our model, we feed 100 definition descriptions generated by GPT-4 for the same diseases (VSD and PDA) and symptoms (Shunt and Hypertrophy) and record the distribution of the resulting F1 scores. As illustrated in Figure~\ref{fig:f1_distribution}, the performance can be affected by varied definition descriptions, which also supports the necessity of our aggregation strategy.

\section{Conclusion}
This study introduces HSDreport, a new benchmark and method designed to revolutionize HSD by integrating the diagnostic gold standard of echocardiography with the accessibility of auscultation. By collecting a novel dataset of 2,275 heart sound samples paired with echocardiography, we set the stage for a more detailed approach to diagnosing heart conditions. The development of our knowledge-aware query-based transformer model marks a significant advancement in the field, effectively harnessing the potential of medically pre-trained models and the nuanced understanding of LLM.

%Our approach not only challenges the traditional boundaries of HSD by moving beyond rigid categorizations but also enhances the interpretative depth and practical relevance of heart sound analysis. The integration of comprehensive echocardiography insights with the practical ease of auscultation represents a substantial leap towards more accurate, robust, and clinically valuable diagnostic tools. The empirical results confirm that our method outstrips existing models, providing superior detection of critical heart abnormalities. This success underscores the importance of blending sophisticated AI techniques with real-world clinical demands to foster advancements that are both scientifically sound and medically valuable.

%Future work will aim to refine these models further, exploring additional integrations of clinical data and expanding the dataset to include a wider array of heart conditions and patient demographics. By continuing to bridge the gap between clinical practice and technological innovation, we can significantly enhance the efficacy and reliability of heart disease diagnostics, ultimately leading to better patient outcomes.
\section*{Limitations}
The dataset used in this study focuses exclusively on pediatric subjects, offering valuable insights specific to this demographic but may not be entirely representative of broader population dynamics. Moreover, our novel dataset is enriched with detailed ultrasonography report data, allowing for potential stratification of severity in abnormal conditions based on numerical indicators. However, the current methodology primarily addresses the presence rather than the severity of these abnormalities, suggesting an avenue for future enhancement. Lastly, due to the multiple inferences required by our approach, the inference speed is slightly slower compared to the baseline model, mainly because we prioritize accuracy over speed.
\section*{Ethical Considerations}
This study was conducted with a strong commitment to ethical standards in medical research, ensuring the protection and confidentiality of participant data and compliance with relevant regulations. Here we outline the ethical considerations addressed in this study.
%\subsection{Informed Consent}
\paragraph{Informed Consent}
All patients, or their legal guardians in the case of minors, provided informed consent for the use of their medical data in this research. Prior to data collection, participants were adequately informed about the nature of the study, the type of data to be collected (heart sounds and ultrasound reports), and the intended use of this data in research. This process was conducted in accordance with the Declaration of Helsinki regarding ethical principles for medical research involving human subjects.
%\subsection{Data Confidentiality and Security}
\paragraph{Data Confidentiality and Security}
Rigorous measures were taken to ensure the confidentiality and security of the data collected. Personal identifiers were removed from all datasets to achieve anonymization. Additionally, all digital data were stored in an encrypted environment to prevent unauthorized access, ensuring that the privacy of the participants was maintained throughout the study.
%\subsection{Compliance with Regulatory Standards}
\paragraph{Compliance with Regulatory Standards}
The study strictly adhered to national laws and regulations concerning medical research and data protection. This adherence was continuously monitored by our legal and ethical advisory board to ensure ongoing compliance throughout the study's duration.
%\subsection{Ethical Review and Oversight}
\paragraph{Ethical Review and Oversight}
The research protocol was thoroughly reviewed and approved by an independent ethics committee. This committee provided continuous oversight and guidance to ensure that all aspects of the study were conducted ethically and that the welfare of the participants was prioritized at all times.
%\subsection{Consideration of Potential Biases}
%\paragraph{Consideration of Potential Biases}
%It is recognized that the data collected from a specific national population, comprising only children, may introduce biases. Consequently, the findings of this study may not be universally applicable across different ethnic or age groups.

\bibliography{custom}

%\appendix

%\section{Example Appendix}
%\label{sec:appendix}

%This is an appendix.

\end{document}